\documentclass[a4paper,10pt,twoside]{cpc-hepnp}

\usepackage{multicol}
\usepackage{graphicx}
\usepackage{booktabs}
\usepackage{amssymb,bm,mathrsfs,bbm,amscd}
\usepackage{slashed}
\usepackage[tbtags]{amsmath}
\usepackage{lastpage}
\usepackage{CJK}

\def\process{$e^+e^-\to \gamma h$}
\begin{document}
\begin{CJK*}{GBK}{song}

\fancyhead[c]{\small Submitted to Chinese Physics C} 
%\fancyfoot[C]{\small 010201-\thepage}

%\footnotetext[0]{Received xx March 2015}

\title{New Physics Searches with \process~at the Higgs Factory}
%在希格斯工厂上利用希格斯粒子与光子联合产生过程进行新物理探测

\author{%
      REN Hong-Yu$^{1,2;1)}$\email{renhy10@mails.tsinghua.edu.cn}%
}
\maketitle

\address{%
$^1$ Department of Physics, Tsinghua University, Beijing 100084, China\\
$^2$ Laboratory for Elementary Particle Physics, Cornell University, Ithaca, NY 14853, USA
}

\begin{abstract}
The Higgs factory is designed for precise measurement of Higgs characters and search for new physics. In this paper we propose that \process~ process could be a useful channel for new physics, which is normally expressed model independently by effective field theory. We calculate the cross section in both the Standard Model and effective field theory approach, and find that the new physics effects of $\gamma h$ have only two degrees of freedom, much fewer than the Higgsstrahlung process. This point could be used to reduce the degeneracies of Wilson coefficients. We also calculated for the first time the $2\sigma$ bounds of $\gamma h$ at the Higgs factory, and prove that $\gamma h$ is more sensitive to some dimension-6 operators than the current experimental data. In the optimistic scenario new physics effects may be observed at the CEPC or FCC-ee after the first couple of years of their run.    
\end{abstract}



%\footnotetext[0]{\hspace*{-3mm}\raisebox{0.3ex}{$\scriptstyle\copyright$}2013
%Chinese Physical Society and the Institute of High Energy Physics
%of the Chinese Academy of Sciences and the Institute
%of Modern Physics of the Chinese Academy of Sciences and IOP Publishing Ltd}%

\begin{multicols}{2}

\section{Introduction}

Following the discovery of the Higgs boson, precise understanding of the nature of this particle is the top priority for particle physics. All measurements of rates involving the Higgs production and decay in the Run 1 of the LHC agree with the predictions of the Standard Model (SM), but statistical uncertainties limit their precision to $10-20$\% level at best. The LHC is expected to ultimately reach a precision of order a few percents, at which point systematic and theoretical issues, such as parton distribution function uncertainties, become a limiting factor. Further improvements in precision are possible at an electron-positron collider with sufficient center-of-mass energy to produce a large sample of Higgs bosons, the so-called "Higgs factory"~\cite{Dawson:2013bba}. Currently, proposals for Higgs factories are being discussed by the physics community, including the CEPC~\cite{CEPC,Ruan:2014xxa}, as well as circular collider designs such as FCC-ee (formerly known as TLEP)~\cite{Gomez-Ceballos:2013zzn,Ruan:2014xxa} and International Linear Collider~\cite{Baer:2013cma}. The physics case for all these machines rests on their ability to test the SM, and search for new physics beyond the SM (BSM), via precision measurements of the Higgs properties. 

The dominant Higgs production process in electron-positron collisions in the energy range relevant for Higgs factories, $\sqrt{s}\sim 225\ldots 350$ GeV, is the Higgsstrahlung process, $e^+e^-\to Zh$. The cross section of this process is expected to be measured with exquisite precision, well below 1\% level, at the Higgs factory. The sensitivity of this measurement to new physics involving the Higgs has been explored by many authors~\cite{Hagiwara:1993sw,Gounaris:1995mx,Kilian:1996wu,GonzalezGarcia:1999fq,Hagiwara:2000tk,Barger:2003rs,Biswal:2005fh,Kile:2007ts,Dutta:2008bh,Contino:2013gna,Amar:2014fpa,Beneke:2014sba,Craig:2014una}. In this paper, we study the Higgs production in association with a photon, \process. In the SM, the leading contribution to the scattering amplitude for this process appears at the one-loop order. As a result, its cross section is strongly suppressed compared to Higgsstrahlung, which occur at tree-level. For this reason, the $\gamma h$ production channel has not received as much attention in the studies of a Higgs factory physics potential so far. However, small SM cross section may offer an advantage in searches for BSM physics, since the BSM effects in the $\gamma h$ channel are expected to produce much larger fractional shifts than in the case of $Zh$. This may compensate for larger statistical uncertainties in the $\gamma h$ rate measurement, resulting in competitive sensitivities to new physics. The goal of this paper is to study this issue quantitatively, in the frameworks for new physics: effective field theory (EFT) approach.

\section{\process ~ in standard model}

The SM cross section for \process~has been computed by several groups~\cite{Barroso:1985et,Abbasabadi:1995rc,Djouadi:1996ws}. We will use the results of Ref.~\cite{Djouadi:1996ws}. The SM cross section with unpolarized beams as a function of the center-of-mass energy is shown in Fig.~\ref{fig:xssm}. The cross section at $\sqrt{s}=250$ GeV, a benchmark energy for Higgs factories, is close to the maximum, about 0.08 fb. This is about 2500 times smaller than the $Zh$ cross section at the same energy, since the $\gamma h$ process is loop-suppressed. Still, with projected luminosities of Higgs factories, a significant number of $\gamma h$ events can be expected. For example, data samples in the 1-10 ab$^{-1}$ range, envisioned in proposals for circular Higgs factories, would contain hundreds of such events. 

\begin{center}
	\includegraphics[width=7cm]{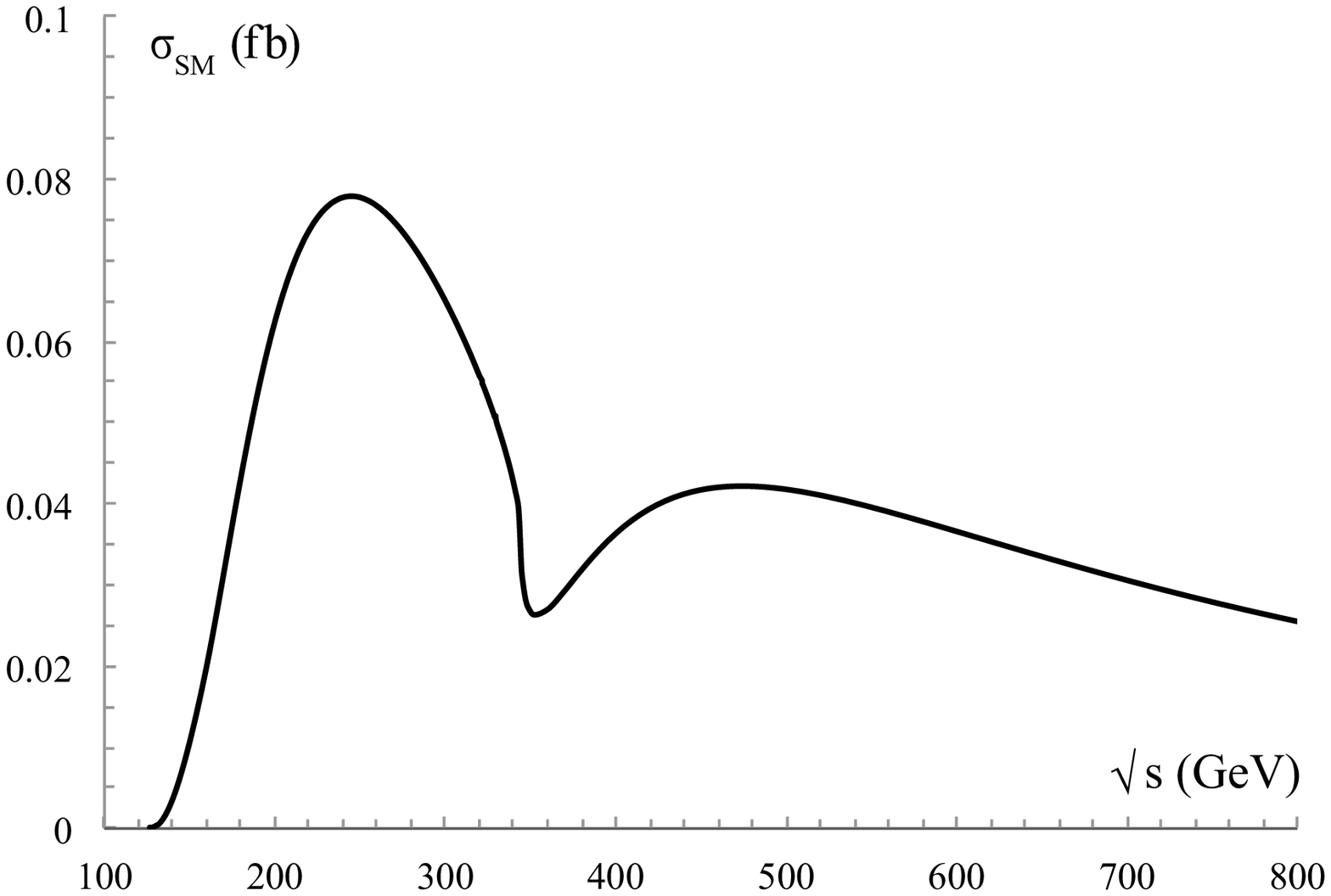}
	\figcaption{\label{fig:xssm}   Figure 1. The cross section of \process~in the Standard Model.}
\end{center}

\begin{center}
	\includegraphics[width=7cm]{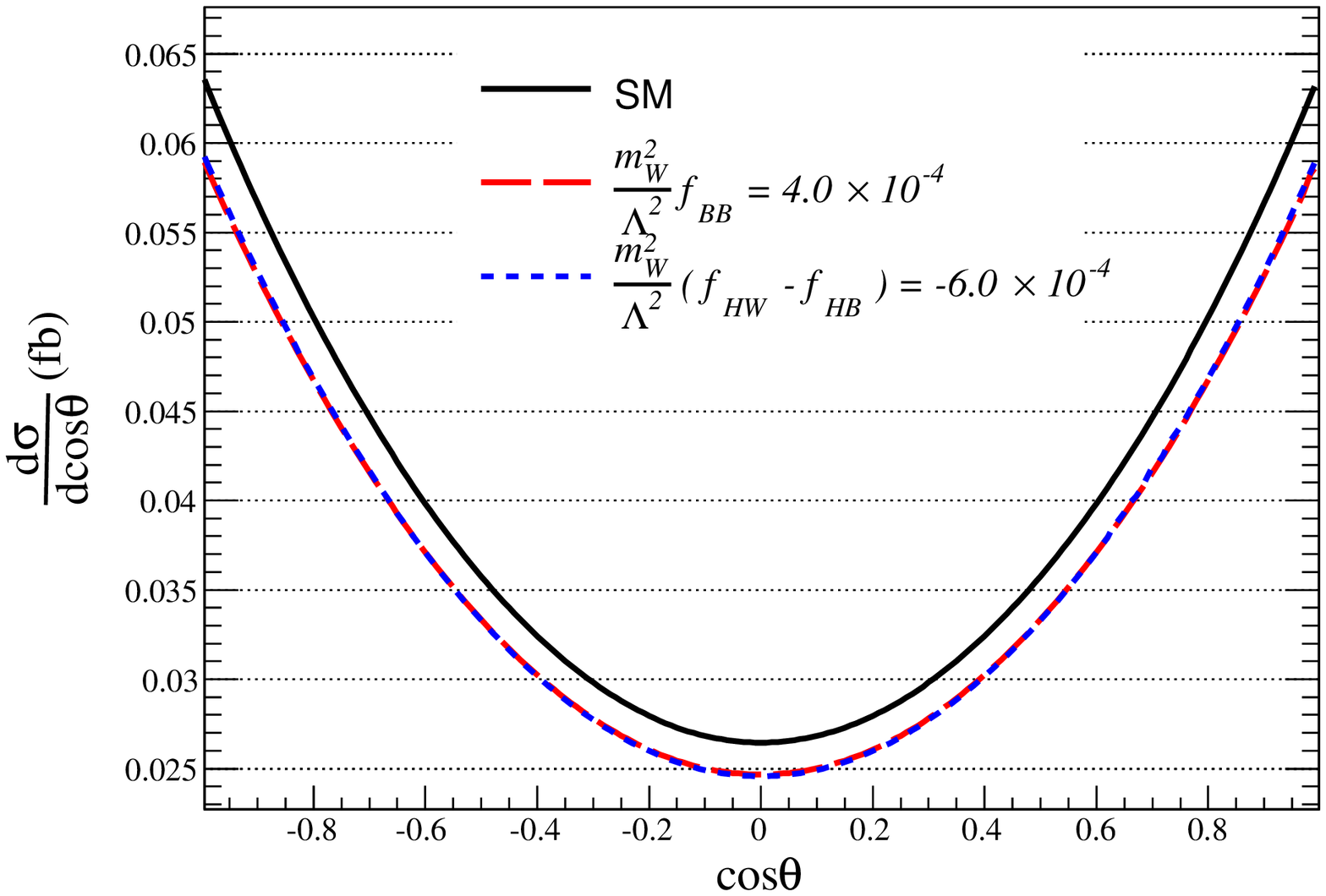}
	\figcaption{\label{fig2}  Figure 2. Photon angular distributions at $\sqrt{s}=250$ GeV, in the SM (black/solid) and in the EFT with two different choices of the dimension-6 new physics operators (red/long-dashed and blue/short-dashed).}
\end{center}

Separation of signal from backgrounds is straightforward. At an $e^+e^-$ collider, the photons produced in association with the Higgs are monoenergetic:
\begin{equation}
E_\gamma=\frac{s-m_h^2}{2\sqrt{s}}.
\end{equation}
At $\sqrt{s}=250$ GeV, this gives a ``spectral line" at $93.75$ GeV. The natural Higgs width being very small, the width of the line is dominated by the detector resolution, which is expected to be $\delta E_\gamma/E_\gamma \approx 1$ \%~\cite{Behnke:2013lya}. This allows for clear separation between the $\gamma h$ line and the much larger $\gamma\gamma$ and $Z\gamma$ lines, at $125.0$ and $108.4$ GeV, respectively. To increase $S/B$ further, one can demand that the Higgs boson be reconstructed, for example as a pair of jets consistent with an invariant mass of $125$ GeV. This requirement will virtually completely eliminate most of the backgrounds, with the left mainly contributed from a $gamma$ and an off-shell $Z$ boson associated production, while the $Z$ boson decays to two jets. The clean environment of the $e^+e^-$ collisions allows for reconstruction of the Higgs with high efficiency in all relevant decay channels. In this study, we will assume that the dominant error on the \process~cross section measurement is statistical, while the significance can be calculated by the relation $S/\sqrt{B}$.

\section{New physics in \process~: Effective field theory}

If new physics appears at a scale $\Lambda\gg \sqrt{s}$, its effects can be described in the language of Effective Field Theory, by adding all possible non-renormalizable operators consistent with gauge and global symmetries of the SM. The leading term in the $\sqrt{s}/\Lambda$ expansion of the Lagrangian contains dimension-6 operators:
\begin{equation}
	{\cal L}_{\rm dim6} = \sum_i \frac{f_i}{\Lambda^2} \, \mathcal{O}_i\,,
\end{equation}
where $f_i$ are dimensionless Wilson coefficients. 

The following dim.-6 operators contribute to the process \process:  
\begin{equation}
	\begin{split}
		\mathcal{O}_{HW}&=ig(D^\mu H)^\dag\sigma^a(D^\nu H)W^a_{\mu\nu},\\
		\mathcal{O}_{HB}&=ig'(D^\mu H)^\dag(D^\nu H)B_{\mu\nu},\\
		\mathcal{O}_{BB}&=g'^2|H|^2 B_{\mu\nu}B^{\mu\nu},\\
		\mathcal{O}_{eW}&=gy_l\bar{L}_L\sigma^a \gamma^{\mu\nu}He_R W^a_{\mu\nu}+{\rm h.c.},\\
		\mathcal{O}_{eB}&=g'y_l\bar{L}_L\gamma^{\mu\nu}He_R B_{\mu\nu}+{\rm h.c.}
	\end{split}
\end{equation}
The last two operators are expected to be Yukawa-suppressed due to chirality flip, and we will not consider them further in this paper. After electroweak symmetry breaking, the first three operators induce $Z\gamma h$ and $\gamma\gamma h$ vertices, leading to a tree-level (but $s/\Lambda^2$-suppressed) contribution to the \process~amplitude. Note that, unlike the $Zh$ process, there is no "contact-interaction" four-particle vertex in this case, since such an interaction is inconsistent with the unbroken $U(1)_{\rm EM}$ gauge symmetry. The new physics contribution to the scattering amplitude is given by
\begin{equation}
	{\cal A}_{\rm EFT} = \sum_{a=+,-}\Lambda^a C^a_{\rm EFT},
\end{equation} 
where 
\begin{equation}
	\Lambda^{\pm}=\bar{v}(p_+)(1\pm\gamma_5)[\slashed\epsilon_\gamma p_\gamma\cdot(p_++p_-) 
		- \slashed p_\gamma\epsilon_\gamma\cdot(p_++p_-)]u(p_-)
\end{equation}
and 
\begin{equation}
	\begin{aligned}
		&C^\pm_{\rm EFT}=-\frac{2 e^2 s_\theta m_W^3}{\Lambda^2} \\
		&\left[\frac{2}{s}f_{BB}+\frac{\lambda^\pm}{8s_\theta^2(1-s_\theta^2)(s-m_Z^2)}(f_{HW}-f_{HB}+8s_\theta^2 f_{BB})\right].
	\end{aligned}
\end{equation}
Here $s_\theta$ is the sine of the Weinberg angle; $p_-$ and $p_+$ are the electron and positron momenta; $s=(p_-+p_+)^2$; and 
\begin{equation}
	\lambda^+=-1+2s_\theta^2,~~~\lambda^-=2s_\theta^2.
\end{equation}
The leading correction to the cross section is due to interference between the SM one-loop amplitude, given in Ref.~\cite{Djouadi:1996ws}, and ${\cal A}_{\rm EFT}$. Numerically, the fractional deviation in the total cross section at $\sqrt{s}=250$ GeV is given by
\begin{equation}
	\label{equ:numeric}
		\begin{aligned}
			\frac{\Delta\sigma(\gamma h)}{\sigma(\gamma h)} \approx 
			&\bigg[ 0.76 (f_{HW} - f_{HB}) - 1.47 f_{BB} \\
			&+ 0.23 (f_{HW}-f_{HB})^2 + 5.63 f_{BB}^2 \\
			&+ 0.59(f_{HW}-f_{HB})f_{BB}\bigg] \Lambda_{\rm TeV}^{-2},
%	\frac{\sigma_{\rm EFT}}{\sigma_{\rm SM}}\approx 1+118(\hat{c}_{HW}-\hat{c}_{HB})-228\hat{c}_{BB},
	\end{aligned}
\end{equation}
where $\Lambda_{\rm TeV}\equiv \Lambda/(1~{\rm TeV})$. For comparison, the fractional shift of the $e^+e^-\to hZ$ cross section at the same energy is~\cite{Craig:2014una}
\begin{equation}
	\label{equ:numerichZ}
	\frac{\Delta\sigma(hZ)}{\sigma(hZ)} \approx \left( 0.05 f_{HW} - 0.005 f_{HB} +0.01 f_{BB} + \ldots\right) \Lambda_{\rm TeV}^{-2},
%	\frac{\sigma_{\rm EFT}}{\sigma_{\rm SM}}\approx 1+118(\hat{c}_{HW}-\hat{c}_{HB})-228\hat{c}_{BB},
\end{equation}
where we omitted the contributions from operators that do not contribute to $\gamma h$. These formulas illustrate the advantage of the $\gamma h$ process mentioned in the Introduction: the SM amplitude is tree-level in $hZ$ and loop-suppressed in $\gamma h$, resulting in a much larger fractional deviation in the cross section in the latter case. 

Estimates of $2\sigma$ exclusion sensitivities at a Higgs factory in the $\gamma h$ channel are listed in Table \ref{tab1}. The estimates assume integrated luminosity of $L_{\rm int}=10$ ab$^{-1}$ at $\sqrt{s}=250$ GeV, corresponding to the FCC-ee projection in~\cite{Ruan:2014xxa}; the sensitivities scale as $L^{-1/2}_{\rm int}$. For these parameters, a sample of about 800 $\gamma h$ events would be collected,  resulting in a cross section measurement with $\delta\sigma/\sigma\approx 20$\% (assuming statistical error dominance and 100\% event reconstruction efficiency). For clarity and ease of comparison among various measurements, the reach for each operator is estimated assuming that all other operators are set to zero. Table \ref{tab1} also list bounds from a global fit to currently available data  \cite{Ellis:2014jta}, such as precision electroweak observables and the Higgs rate measurements at the LHC. For two of the three relevant operators, $\mathcal{O}_{HB}$ and $\mathcal{O}_{HW}$, the $\sigma(\gamma h)$ measurement at the Higgs factory will probe scales exceeding the current bounds. The third operator, $\mathcal{O}_{BB}$, is already very well constrained by the measurement of Br($h\to\gamma\gamma$) at the LHC, where the competing SM amplitude only appears at the one-loop order. In this case, neither $\gamma h$ nor $Zh$ channel could perform better than current data. However, it should be emphasized that this is so only as long as the operators are turned on one-by-one; the LHC bound on $\hat{c}_{BB}$ can be significantly relaxed if other operators, for example $\mathcal{O}_{GG}=|H|^2G^a_{\mu\nu}G^{a\mu\nu}$, are present. 
The measurement of $\gamma h$ cross section at the Higgs factory will allow to resolve such ambiguities.

The operators that contribute to \process~will also modify the $Zh$ cross section. For comparison, the sensitivities of this measurement is also listed in Table \ref{tab1}. In all cases, we assumed that statistical errors dominate, and used the same benchmark value of 10 ab$^{-1}$ for integrated luminosity. (As long as the precision is statistics-limited, all estimates scale as $L^{-1/2}_{\rm int}$, so that statements concerning the relative power of various measurements remain valid.) For all three operators, $\sigma(Zh)$ measurements have somewhat higher reach compared to the $\sigma(\gamma h)$ measurement. Still, including $\sigma(\gamma h)$ in a global fit should give a meaningful improvement in sensitivity to new physics. 

\end{multicols}

\begin{center}
\tabcaption{ \label{tab1}  Current 95\% CL bounds (2nd column) and future Higgs factory $2\sigma$ exclusion sensitivities (3rd-4th columns) on the coefficients of the dim.-6 operators that contribute to \process. Here $\hat{c}_i=m_W^2 f_i/\Lambda^2$.The current bounds are taken from Ref.~\cite{Ellis:2014jta}. Higgs factory estimates assume that statistical uncertainties dominate. The main background of $\gamma h$ is included while that of $Zh$ is not, because the huge cross section of $Zh$ can supress the effects of background.}
\footnotesize
\begin{tabular*}{150mm}{@{\extracolsep{\fill}}cccc}\hline
Coefficients & Current Bound & $\sigma(\gamma h)$ & $\sigma(Zh)$\\
\hline
$\hat{c}_{HW}$ & $(-0.042, ~0.008)$ & $(-0.0050,~0.0033)$ & $(-1.8,~1.8)\times 10^{-4}$\\
$\hat{c}_{HB}$ & $(-0.053, ~0.044)$ & $(-0.0033,~0.0050)$ & $(-1.8,~1.8)\times 10^{-3}$\\
$\hat{c}_{BB}$ & $(-4.0, ~2.3)\times 10^{-4}$ & $(-0.0012,~0.0028)$ & $(-9,~9)\times 10^{-4}$\\
\bottomrule
\end{tabular*}
\end{center}

\begin{multicols}{2}

In general, angular distributions of final-state particles may contain additional information allowing for better discrimination between SM and new physics, and also, should a new physics effect be observed, between various possible combinations of dim.-6 operators. Unfortunately, in the case of~\process, no new information is contained in the photon angular distribution,  as is clear from Fig.~\ref{fig:xssm} (right panel). There may be additional information in angular correlations between $\gamma$ and the Higgs decay products; we defer a study of such correlations for future work.

So far, we've considered bounds in the situation where a single dimension-6 operator is assumed to be dominant. More generally, each observable constrains a particular linear combinations of operators, leaving a subspace in the operator coefficient space unconstrained. For example, if described by effective field theory, the new physics in $Zh$ has about 10 degrees of freedom\cite{Craig:2014una} but only one observable. So we need more observables to reduce, even eliminate such degeneracies. $\gamma h$ could be one of such observables. From Eq. \ref{equ:numeric} we can see that the cross section of $\gamma h$ has only two degrees of freedom, $f_{HW}-f_{HB}$ and $f_{BB}$, while the latter one has been constrained strictly by the current data. This means if new physics effect is observed througn $\gamma h$ in future, we can almost be certain that it comes from the $\mathcal{O}_{HW}$ or $\mathcal{O}_{HB}$. This is the advantage of $\gamma h$ compared to $Zh$, and is why we claim that $\gamma h$ is quite valuable although it is less sensitive than $Zh$. 

\begin{center}
	\includegraphics[width=7cm]{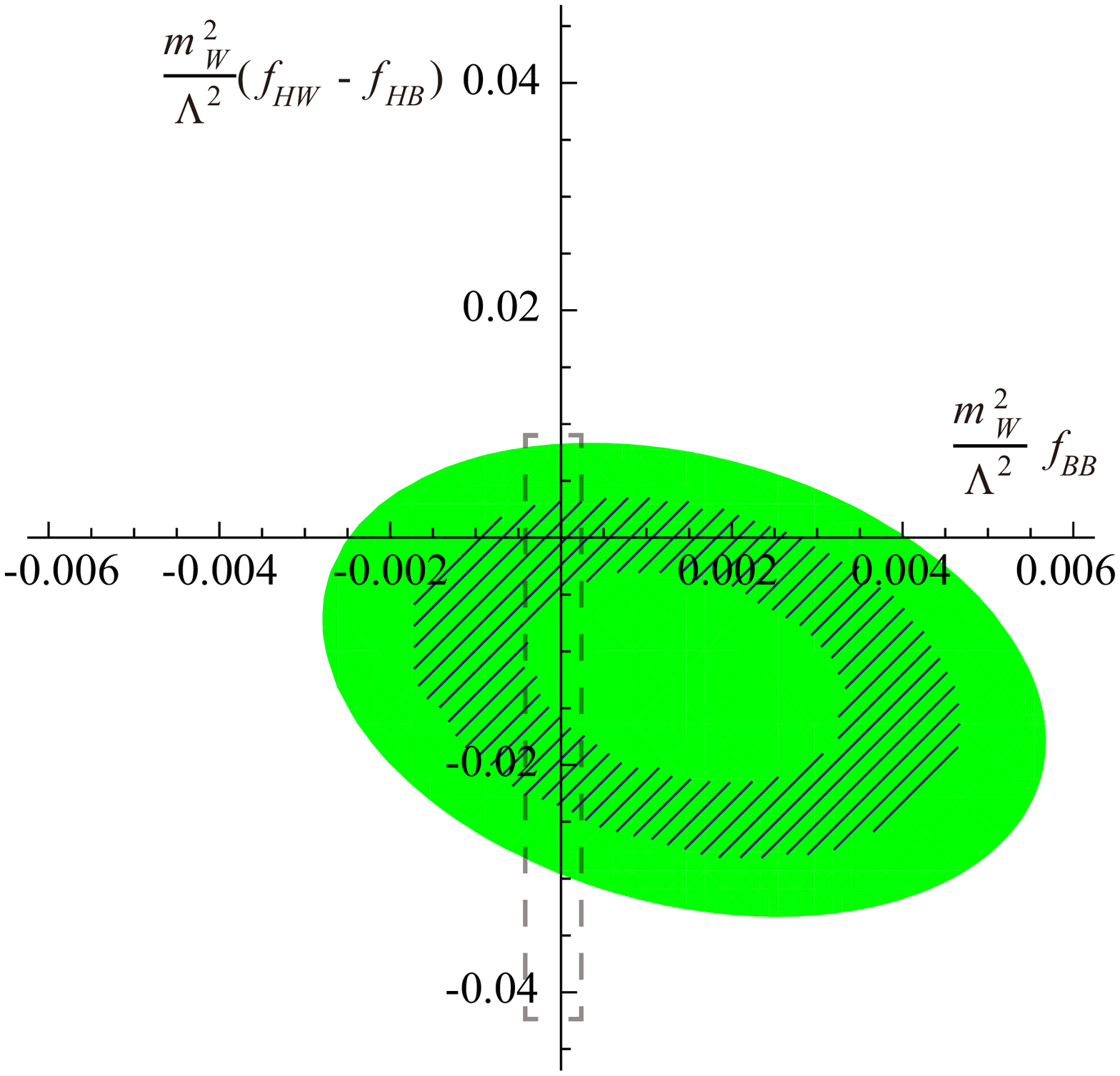}
	\figcaption{\label{fig:2sigma} The $2\sigma$ sensitivities of $\gamma h$ at the Higgs factory with different integrated luminosities. The bounds are obtained through a two-parameter analysis. The shaded region is where the effective operators are beyond the sensitivity of $\gamma h$ at the integrated luminosity of $10~{\rm ab}^{-1}$, which could be provided by FCC-ee in about 5 years. The green region corresponds to 1 ab$^{-1}$, which could be provided by CEPC in about 2 years\cite{Ruan:2014xxa}. The dashed grey lines are the current $2\sigma$ bounds obtained with single-parameter analysis.}
\end{center}

We can also implentment two-parameter analysis on the 95\% CL bounds of $\gamma h$, as shown in Figure \ref{fig:2sigma}. From this figure it is clear that $\gamma h$ can be helpful to measure or give bounds to the Wilson coefficients $f_{HW}-f_{HB}$ specially. If there are new physics effects within the sensitivity of $\gamma h$, it is hopeful that we can see them at the first few years of the run of CEPC or FCC-ee. Otherwise we could give limits on $f_{HW}-f_{HB}$, and these limits can be applied to $Zh$ or other processes to extract more information about new physics effects. 

We close this section with a comment of a technical nature. Numerical SM predictions of cross sections such as $\sigma(\gamma h)$ depend on the values of the electroweak gauge couplings and the Higgs vacuum expectation value, which are inferred from the three most precisely measured electroweak observables, currently $M_Z$, $\alpha$, and $G_F$ (from muon lifetime). New physics can contribute to these observables, producing a shift between the inferred and the true values of these parameters. In general, such shifts contribute to the deviation of cross sections from their SM values. For example, in the case of $\sigma(hZ)$, the contribution of such coupling shifts is of the same order as the direct contribution of the dim.-6 operators, and both need to be taken into account for consistency~\cite{Craig:2014una}. However, in the case of $\sigma(\gamma h)$, where the leading SM amplitude is one-loop, the correction of the scattering amplitude due to the coupling shifts is of order $\frac{1}{16\pi^2}\frac{s}{\Lambda}$, whereas the direct contribution of dim.-6 operators is of order $\frac{s}{\Lambda^2}$. The additional loop factor in the coupling shift correction renders it negligible, and we do not include this effect in our analysis.       

\section{Conclusion}

The Higgs factories, normally known as CPEC, FCC-ee and ILC, are designed to study the Higgs couplings with other particles precisely, by producing a large number of Higgs bosons mainly through the Higgsstrahlung process, $e^+e^-\to Zh$. The Higgsstrahlung is commonly believed to be one of the most precise processes for measuring the Higgs couplings, and it can be very sensitive to new physics effects. However, if described by effective field theory, the Higgsstrahlung has too many new physics degrees of freedom, and not enough observables. This may cause degeneracies and "blind spots" where new physics effects escape from the reach of the detectors. In this paper, we propose a new idea, the $e^+e^-\to \gamma h$ channel as a supplement of the Higgsstrahlung to detect the new physics effects. The advantage of $\gamma h$ is that it has only two degrees of freedom and also a good sensitivity, compared to the current data. With the help of $\gamma h$, we can extract the information on $f_{HW}-f_{HB}$ and this will help us know the features of new physics better. In our future work, we will also study the Higgs decay processes $h\to \gamma \gamma$ and $h\to Z\gamma$. These two decay modes have the same degrees of freedom as $\gamma h$ and are believed to be sensitive to BSM effects too. The $h\to \gamma\gamma$ decay has been studied in \cite{Dawson:2013bba,Ruan:2014xxa} but $h\to Z\gamma$ has not. These two channels may be valuable in reducing the degenereacies of Wilson coefficients, playing as a cross check of the $Zh$ and $\gamma h$ results, and is worth a exhaustive analysis.

\mbox{}\\

\acknowledgments{I am very grateful for the guidance of Maxim Perelstein, and the conversations with Yu-Ping Kuang and Ling-Hao Xia. H.-Y. Ren is supported by the National Natural Science Foundation of China under Grants No. 11275102, and Tsinghua Scholarship for Overseas Graduate Studies.}

\end{multicols}

\vspace{10mm}

\vspace{-1mm}
\centerline{\rule{80mm}{0.1pt}}
\vspace{2mm}

\begin{multicols}{2}

%\bibliographystyle{unsrt}
%\bibliography{unsrt}

\end{multicols}

\clearpage

\end{CJK*}
\end{document}